\documentclass[conference]{IEEEtran}
\usepackage{cite}      % Written by Donald Arseneau
\usepackage{graphicx}  % Written by David Carlisle and Sebastian Rahtz

\begin{document}

% paper title
\title{Is Spiking Logic the Route to Memristor-Based Computers?}

% author names and affiliations
% use a multiple column layout for up to three different
% affiliations
\author{\authorblockN{Ella Gale}
\authorblockA{Unconventional Computing Centre\\
Bristol Robotics Laboratory\\
University of the West of England\\
Bristol, UK BS16 1QY\\
Email: ella.gale@uwe.ac.uk}
\and
\authorblockN{Ben de Lacy Costello}
\authorblockA{Unconventional Computing Centre\\
Faculty of Applied Sciences\\
University of the West of England\\
Bristol, UK BS16 1QY\\
Email: ben.delacycostello@uwe.ac.uk}
\and
\authorblockN{Andrew Adamatzky}
\authorblockA{Unconventional Computing Centre\\
Computer Science and Creative Technology\\
University of the West of England\\
Bristol, UK BS16 1QY\\
Email: andrew.adamatkzy@uwe.ac.uk}}

% avoiding spaces at the end of the author lines is not a problem with
% conference papers because we don't use \thanks or \IEEEmembership

% for over three affiliations, or if they all won't fit within the width
% of the page, use this alternative format:
% 
%\author{\authorblockN{Michael Shell\authorrefmark{1},
%Homer Simpson\authorrefmark{2},
%James Kirk\authorrefmark{3}, 
%\Montgomery Scott\authorrefmark{3} and
%Eldon Tyrell\authorrefmark{4}}
%\authorblockA{\authorrefmark{1}School of Electrical and Computer Engineering\\
%Georgia Institute of Technology,
%Atlanta, Georgia 30332--0250\\ Email: mshell@ece.gatech.edu}
%\authorblockA{\authorrefmark{2}Twentieth Century Fox, Springfield, USA\\
%Email: homer@thesimpsons.com}
%\authorblockA{\authorrefmark{3}Starfleet Academy, San Francisco, California 96678-2391\\
%Telephone: (800) 555--1212, Fax: (888) 555--1212}
%\authorblockA{\authorrefmark{4}Tyrell Inc., 123 Replicant Street, Los Angeles, California 90210--4321}}

% use only for invited papers
\specialpapernotice{(Invited Paper)}

% make the title area
\maketitle

\begin{abstract}
Memristors have been suggested as a novel route to neuromorphic computing based on the similarity between neurons (synapses and ion pumps) and memristors. The D.C. action of the memristor is a current spike, which we think will be fruitful for building memristor computers. In this paper, we introduce 4 different logical assignations to implement sequential logic in the memristor and introduce the physical rules, summation, `bounce-back', directionality and `diminishing returns', elucidated from our investigations. We then demonstrate how memristor sequential logic works by instantiating a NOT gate, an AND gate and a Full Adder with a single memristor. The Full Adder makes use of the memristor's memory to add three binary values together and outputs the value, the carry digit and even the order they were input in. 

\end{abstract}

% no keywords

% For peer review papers, you can put extra information on the cover
% page as needed:
% \begin{center} \bfseries EDICS Category: 3-BBND \end{center}
%
% for peerreview papers, inserts a page break and creates the second title.
% Will be ignored for other modes.
\IEEEpeerreviewmaketitle

\section{Introduction}
% no \PARstart
Memristors have been compared to both neurons~\cite{247,248} and synapses~\cite{71,216,236} and have widely been anticipated as a useful route towards neuromorphic (brain-like) computing due to the memristor's ability to hold a memory or state~\cite{14}. Although the memristor was predicted to exist based on symmetry concerns~\cite{14} real world analogues were not recognised to exist until 2008~\cite{15}, even though they were made before~\cite{119}. The memristor is commonly considered as an A.C. element~\cite{276}, however the D.C. response of the memristor is highly interesting because memristors possess spike-like dynamics~\cite{SpcJ} which have been shown to combine in spike-train-like ways in memristor networks~\cite{c0c}. Furthermore, the interactions of these spikes, which can be considered the short-term memory of the memristor, can be used with a novel sequential logic approach which can be used to made simple logic circuits~\cite{P0c}.

The advantages of using spike interactions are many-fold. The memristor switching itself can be slow~\cite{260} but the spikes can interact much faster, the output of which is `held' in the short-term memory of the memristor, which gives rise to, if not faster processing, more complex operations within a given time-frame than is usually the case in standard electronics.

When designing devices to compute with a certain logic, we have some freedom in how the logical `1's and `0's are assigned. For these devices we shall take the voltage inputs as the logical input, with the current values as the logical output, where it is understood that some processing (via a memristor or other circuitry) is required in order to instantiate the logical circuit. Within this assumption, it is possible to compute operations of a surprisingly high complexity with just a single memristor, and, it is suggested, that with this approach and the conversion circuitry, a useful approach to computation.

In this paper we will examine in more detail the physics of the memristors and the physical rules in order to understand the operation of sequential logic and present some examples of the high-level computation that a single memristor can obtain.

\section{Methodology}

\subsection{Sequential Logic}

We shall make use of sequential logic, as implemented in~\cite{P0c}, which works with  the spike interactions seen in the memristor. Memristor sequential logic allows the computation through time by storing a state and allowing it to interact with the input; thus a one terminal device can do two-input (or higher) logical operations, if we are willing to wait for the output. 

\begin{figure}
\centering
\includegraphics[width=2.5in]{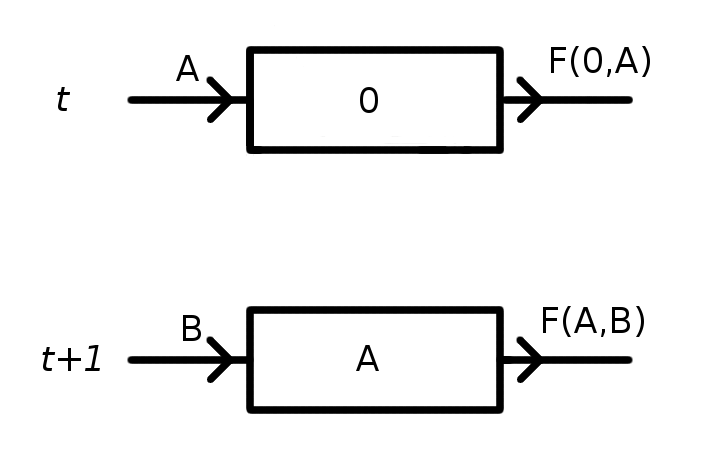}
% where an .eps filename suffix will be assumed under latex, 
% and a .pdf suffix will be assumed for pdflatex
\caption{Sequential Logic. The output of the memristor is a function of its state, as shown in the box, and the input. As the state is stored for the duration of the short-term memory logical values can be combined if they are input sequentially.}
\label{fig:SeqLogic}
\end{figure}

As shown in figure~\ref{fig:SeqLogic} the memristor's state is stored in its short-term memory and the current output to a voltage change is actually a function of its zeroed/null state and the input. Sequential logic makes use of the memristor's short-term memory to store the first bit, $A$, of an operation before the transmission of the second bit $B$. The output at time, $t_A$, is a function of $A$ and the memristor's starting state (which is $\emptyset$ if the device has been properly zeroed), given by $f(A,\emptyset)$. At time $t_B$ (where $t_B$ is one measurement step after $t_A$) the output would be $f(B,A)$. The response step, $t_1$ is measured one measurement step after $t_b$. Thus, this voltage data is input at $t_A$ and $t_B$ and measured at $t_1, t_2$ and so on where $t_a < t_b < t_1 < t_2$. 

\subsection{Experimental}

Memristors were fabricated as in~\cite{260} using the TiO$_2$ sol-gel as described in~\cite{M0} and were measured using a Keithley electrometer, with a set time-step of 0.1s, which gives an actual output of 0.16s. After each logical test, the memristor was left for 40 timesteps ($\sim40$s) to lose its short-term memory, i.e. reset to the null state. All presented results are experimental data.

There are two variables we can utilise when assigning logical values: the magnitude, as represented by $M$ for a high magnitude and $m$ for a low magnitude; and the sign, as represented by a $+$ for positive and $-$ for negative. The 4 different logical assignations that can be applied using these values is shown in table~\ref{tab:Logic}. To implement logical operations, voltage spikes are applied for one time-step and the response recorded at the same frequency. In between logical operations, the devices were left for longer than the equilibration time ($\tau_{\infty}$ in~\cite{SpcJ} which is around 3.5s) to zero the memristor by removing its short term memory. 

% An example of a floating table. Note that, for IEEE style tables, the 
% \caption command should come BEFORE the table. Table text will default to
% \footnotesize as IEEE normally uses this smaller font for tables.
% The \label must come after \caption as always.
%
\begin{table}
%% increase table row spacing, adjust to taste
%\renewcommand{\arraystretch}{1.3}
\caption{Four different methods of implementing logical `1' and `0' with memristor spikes: $M$ refers to a high magnitude voltage, $m$ to a low magnitude voltage and `+' and `-' refer to its polarity.}
\label{tab:Logic}
\begin{center}
%% Some packages, such as MDW tools, offer better commands for making tables
%% than the plain LaTeX2e tabular which is used here.
\begin{tabular}{|c||c|c|c|c|}
\hline
Logical 	& Magnitude	& Polarity	& Mixed 	& Mixed 	\\
value		& Logic		& Logic& Logic 1	& Logic 2	\\
\hline
One 		& M 			& +		& +M		&  -M	\\
Zero 		& m 			& - 	& -m		& +m\\
\hline
\end{tabular}
\end{center}
\end{table}

\section{Elucidated Rules}

Changing the values of $M$ and $m$ can allow the results to be tuned or balanced against the effect of polarity, but in this paper we shall just deal with qualitative examples. From investigation of these systems, we have elucidated the following physical rules for the system.

\subsection{Directionality}

The memristor naturally implements Implication (as first invented by Bertrand Russell and observed in~\cite{242}). The memristor is directional: e.g. The response at $t_1$, for $A \rightarrow B$ does not equal the output ($t_1$) for $B \rightarrow A$. The cause for this is that the memristor responds to the difference in voltage. This naturally allows memristor-based sequential logic to compute implication logic as Implication, IMP or $\rightarrow$, requires that $0 \rightarrow 1 \neq 1 \rightarrow 0$ and thus the order in which the two values are input has a meaning. Naturally, sequential logic, as it separates the values in time, implements this ordering. Note that sequential logic is a scheme for how the memristor can enact logical operations, Implication is an example of a logical operation, and implication logic is the name for the logical set of [IMP, FALSE] required for functionally-complete computation.

\subsection{`Summation' via Energy Conservation}

If the logical `1' is taken as being a high voltage, i.e. $M$ instead of $m$, then more energy is imparted to the system from the logical combinations like [1,1] compared to [0,0]. This approach can allow the creation of memristor based time-limited summators of use in leaky integrate and fire neurons.

\subsection{`Bounceback'}

The application of a voltage spike produces a resultant current spike in the direction of the difference between the starting voltage and the ending voltage, e.g. the first voltage change $V_0 \rightarrow V_A$ causes a positive current reponse, $+i_A$, if $V_A$ is positive, and negative, ,$-i_A$, if $V_A$ is negative. If the system is then returned to zero, there is a smaller current spike of the opposite polarity, i.e. $-i_0$ and $+i_0$ respectively for the two examples mentioned above. If several spikes are input before returning to zero, i.e. a sequence of $[V_0, V_A, V_A, V_0]$ the current spike is larger than would be the case for $[V_0, V_A, V_0]$, although not twice as large due to losses in the system.

\subsection{`Diminishing Returns'}

The effect of additional spikes of the same size and polarity, occurring within the window of the memristor's short term memory decreases. A similar effect is seen with changing polarity, in that changing polarity can cause a larger response (than not) but this response is smaller with successive voltages, i.e. the response spike to $[V_0, +V_A, -V_A, +V_A, -V_A]$ is smaller than $[V_0, +V_A, -V_A]$.

\section{Examples of Logical Systems}

Knowledge of these rules and effects allows us to design logical computation systems which perform a surprising amount of computation with only a single memristor. We have found that, in these schemes that the summation effect is important in magnitude logic, the `bounceback' effect is more relevant in polarity logic (although both affect the outcome). As these can be balanced and set in opposition to each other, the richest effects came from using the mixed logics (as presented in~\ref{tab:Logic}: we will now present a few examples. 

\subsection{Inverter}

\begin{figure}
\centering
\includegraphics[width=3.5in]{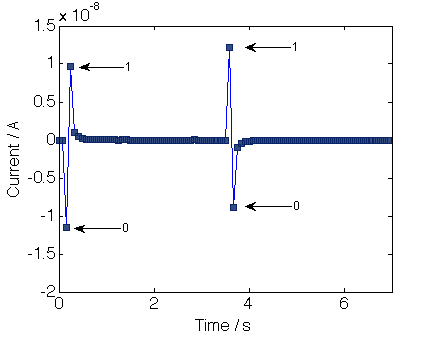}
% where an .eps filename suffix will be assumed under latex, 
% and a .pdf suffix will be assumed for pdflatex
\caption{NOT Gate Implementation}
\label{fig:NOT}
\end{figure}

Using polarity logic and the `bounce-back' effect, it is an easy thing to build an inverter as shown in figure~\ref{fig:NOT}. Because the response spike is in the opposite direction, taking that as the result of the operation switches from 1 to 0 (and vice versa) and can be viewed as performing the NOT operation on the input.

\subsection{AND Gate}

\begin{figure}
\centering
\includegraphics[width=3.5in]{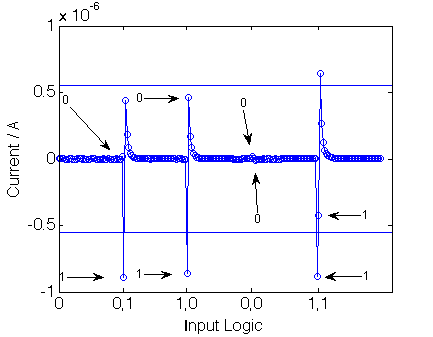}
% where an .eps filename suffix will be assumed under latex, 
% and a .pdf suffix will be assumed for pdflatex
\caption{AND Gate Implementation}
\label{fig:AND}
\end{figure}

An example of an AND gate is shown in figure~\ref{fig:AND}, this example uses mixed logic 2 with a $M$ of -0.5V and a $m$ of +0.001V. If we take the response output as `1' if current over a threshold (in this case, 0.55$\mu A$) is seen, the device implements an AND gate (this is still the case if we choose to limit ourselves to only the value of the $t_1$ response spike). Due to the summation effect, the amount of energy in the [1,1] system is larger than the [0,1], [1,0] and [0,0] parts of the truth-table, and this causes a larger `bounceback' response which can be measured in the positive current response. 

Were we to limit ourselves to the negative current part of the device response, the magnitude of the output picks out an inclusive OR operation, in that the only parts of the truth table that have a response over the threshold are those that contain a `1' (because these spikes depend on a `1' input). Although this response is trivial, it is information that can be usefully used with the correct output circuitry.

%\begin{figure}
%\centering
%\includegraphics[width=2.5in]{myfigure}
% where an .eps filename suffix will be assumed under latex, 
% and a .pdf suffix will be assumed for pdflatex
%\caption{Simulation Results}
%\label{fig_sim}
%\end{figure}

\subsection{Towards a Full-Adder}

\begin{figure}
\centering
\includegraphics[width=3.5in]{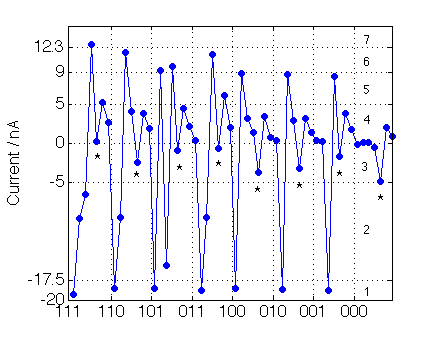}
% where an .eps filename suffix will be assumed under latex, 
% and a .pdf suffix will be assumed for pdflatex
\caption{An attempt at a full adder using mixed logic 2. The first three input bits are the logical inputs, the system has one timestep to respond (\~1s) before a read spike is sent in, as marked by an *. The numbers of the ranges correspond to the list.}
\label{fig:FullAdder}
\end{figure}

It is possible to compute an unconventional instantiation of full-adder, as shown in figure~\ref{fig:FullAdder} (admitting that we require a voltage spike to current spike conversion). The two input and carry bits are input as a series of spikes using mixed logic 2 with input `1' represented by -0.5V and input `0' represented by +0.001V. The input sequence is [A,B,C,1,2,3,4], with the logic input at $t_A-t_C$, the response spike recorded at $t_1$, an extra read voltage of -0.15V input at $t_2$. This gate requires a clock to operate. Figure~\ref{fig:FullAdder} shows the response of the memristor to this scheme, for the three inputs of a full adder, the read spike at $t_2$ is marked with an * to make it easier to understand, and the data of the memristor losing its short-term memory is not shown. 

From this set-up the following things can be deduced from knowing the maximum positive and negative current spikes within 4 time-steps of an input (although this requirement need not be too stringent if we have a way of recording the maximum current within the ranges in between zeroing the system, which we can do with knowledge of the read pulse clock). 

The resulting information from the current is thus:
\begin{enumerate}
\item if a negative current is recorded in the range -17.5 to -20nA: we have had a `1' input into the system
\item if a negative current is recorded in the range -5 to -17.5nA: we have a carry bit from the operation
\item if a negative current is recorded in the range 0 to -5nA: we have had a zero in the system (this is redundant information)
\item if the maximum positive current is recorded in the range 0 to +5nA: the result of the calculation is `0'
\item if the maximum positive current is recorded in the range +5 to +9nA: the result of the calculation is `1'
\item if the maximum positive current is recorded in the range +9 to +12.3: the result is `2' (or `1' for the carry bit, `0' for the summation bit) 
\item if the maximum positive current is recorded over 12.5nA: the result is `3' (or `1' for both the carry and summation bit in binary logic). 
\end{enumerate}

The output in the negative is purely a result of the input voltages to the system. The positive system includes the `bounceback', and the summation effect as probed by the read voltage which gives threasholded values of the memristor's state. 

With switches, it would be possible to send on the logical result as binary. Region two of the plot encodes the carry bit for the operation, because only if there are two $-M$ spikes (which encode `1') within 3 time-steps of each other we will see a current response in that range. The summation bit is not encoded in as direct a manner, the maximum of the positive currents encodes the numerical sum, and so the summation bit for the value 3 is in a different place to that for the value 1. If we only require knowledge of the carry and summation bit, we can do without the read voltage and corresponding spikes. Changing the values of $M$ and $m$ can tune the effect and might allow us to change the relative values of the output spikes.

%\begin{figure}
%\centering
%\includegraphics[width=2.5in]{myfigure}
% where an .eps filename suffix will be assumed under latex, 
% and a .pdf suffix will be assumed for pdflatex
%\caption{Simulation Results}
%\label{fig_sim}
%\end{figure}

\section{Conclusions}

In this short paper we have summarised the physical aspects and causes of the spiking interactions observed in a large number of experimental tests and demonstrated some of the interactions via the creation of NOT, AND and Full Adder gates. These gates are not cascadable, because the output is a different form to this input, nonetheless, the high degree of functionality of a single memristor suggests that solving this problem will a worthwhile endeavour. 

In terms of logical operational complexity, we are not sure if a Full Adder is the limit for a single memristor. The example shown here takes in 3 bits of information and the output includes: the sum, the value of the carry bit, whether the input includes a 1, whether the input includes a 0 and, from the precise value of the spikes at $t_2$, and from a more precise threasholding over the outputs, we can learn where the zero is in the 2 input sequences ([011,101,110]) and where the `1' is in the 1 input sequences ([001,010,100]). These last two points are interesting as it suggests that the memristor Full Adder shown here does not destroy information by the operation, however the gate is not-reversible (as this would require the ability to run time backwards to reverse the physics!). Thus we suggest that with clever design, the memristor can be made to compute more information. As the memristor has to be zeroed, and this takes time, we would want the memristor to do the maximum amount of processing, which suggests that a processor built out of memristors would have a lower clock speed but may compute more bits of information each cycle.

The use of memristor summation approaches in the full adder scheme is similar to how neurons work. For example three `1' inputs received one after the other causes the largest response spike and the only positive $t_2$ spike, either of these outputs could be linked to a threasholded switch which could release a current or voltage spike and thus act like a leaky integrate and fire neuron. The diminishing returns effect could enforce a refractory period. As neurons work by converting a rate-coded spiking voltage to a current spike at the synapse and then to a voltage spike, all of which can be considered transmission of a logical `1', the memristor with its action whereby input and output are current and voltage, could be ideally suited to neuromorphic computing.

\bibliographystyle{IEEEtran.bst}

\maketitle
%\bibstyle{IEEEtran.bst}
% argument is your BibTeX string definitions and bibliography database(s)
\bibliography{IEEEabrv,./UWELit}

\begin{thebibliography}{10}
\providecommand{\url}[1]{#1}
\csname url@rmstyle\endcsname
\providecommand{\newblock}{\relax}
\providecommand{\bibinfo}[2]{#2}
\providecommand\BIBentrySTDinterwordspacing{\spaceskip=0pt\relax}
\providecommand\BIBentryALTinterwordstretchfactor{4}
\providecommand\BIBentryALTinterwordspacing{\spaceskip=\fontdimen2\font plus
\BIBentryALTinterwordstretchfactor\fontdimen3\font minus
  \fontdimen4\font\relax}
\providecommand\BIBforeignlanguage[2]{{%
\expandafter\ifx\csname l@#1\endcsname\relax
\typeout{** WARNING: IEEEtran.bst: No hyphenation pattern has been}%
\typeout{** loaded for the language `#1'. Using the pattern for}%
\typeout{** the default language instead.}%
\else
\language=\csname l@#1\endcsname
\fi
#2}}

\bibitem{247}
L.~Chua, V.~Sbitnev, and H.~Kim, ``Hodgkin-huxley axon is made of memristors,''
  \emph{International Journal of Bifurcation and Chaos}, vol.~22, p. 1230011
  (48pp), 2012.

\bibitem{248}
------, ``Neurons are poised near the edge of chaos,'' \emph{International
  Journal of Bifurcation and Chaos}, vol.~11, p. 1250098 (49pp), 2012.

\bibitem{71}
S.~H. Jo, T.~Chang, I.~Ebong, B.~B. Bhadviya, P.~Mazumder, and W.~Lu,
  ``Nanoscale memristor device as a synapse in neuromorphic systems,''
  \emph{Nanoletters}, vol.~10, pp. 1297--1301, 2010.

\bibitem{216}
G.~Howard, E.~Gale, L.~Bull, B.~de~Lacy~Costello, and A.~Adamatzky, ``Towards
  evolving spiking networks with memristive synapses,'' in \emph{Artificial
  Life (ALIFE), 2011 IEEE Symposium on}, april 2011, pp. 14 --21.

\bibitem{236}
C.~T. Themistoklis~Prodromakis and L.~Chua, ``Two centuries of memristors,''
  \emph{Nature Materials}, vol.~11, pp. 478--481, 2012.

\bibitem{14}
L.~O. Chua, ``Memristor - the missing circuit element,'' \emph{IEEE Trans.
  Circuit Theory}, vol.~18, pp. 507--519, 1971.

\bibitem{15}
D.~B. Strukov, G.~S. Snider, D.~R. Stewart, and R.~S. Williams, ``The missing
  memristor found,'' \emph{Nature}, vol. 453, pp. 80--83, 2008.

\bibitem{119}
L.~Chua, ``Resistance switching memories are memristors,'' \emph{Applied
  Physics A: Materials Science \& Processing}, pp. 765--782, 2011.

\bibitem{276}
C.~L. S.~Grimes and O.~Martinsen, ``Memristive properties of human sweat
  ducts,'' \emph{World Congress on Medical Physics and Biomedial Engineering},
  vol. 25/7, pp. 696--698, 2009.

\bibitem{SpcJ}
E.~Gale, B.~de~Lacy~Costello, and A.~Adamatzky, ``Observation, characterization
  and modeling of memristor current spikes,'' \emph{Appl. Math. Inf. Sci.},
  vol.~7, pp. 1395--1403, 4,July 2013.

\bibitem{c0c}
------, ``Observations of bursting spike patterns in simple memristor
  circuits,'' in \emph{Submitted}.

\bibitem{P0c}
------, ``Boolean logic gates from a single memristor via low-level sequential
  logic,'' in \emph{Submitted}.

\bibitem{260}
E.~Gale, D.~Pearson, S.~Kitson, A.~Adamatzky, and B.~de~Lacy~Costello,
  ``Aluminium electrodes effect the operation of titanium oxide sol-gel
  memristors.'' \emph{arXiv:1106.6293v1}, 2011.

\bibitem{M0}
E.~Gale, R.~Mayne, A.~Adamatzky, and B.~de~Lacy~Costello, ``Drop-coated
  titanium dioxide memristors,'' \emph{Materials Chemistry and Physics}, vol.
  143, pp. 524--529, January 2014.

\bibitem{242}
J.~Borghetti, G.~D. Snider, P.~J. Kuekes, J.~J. Yang, D.~R. Stewart, and R.~S.
  Williams, ```memristive' switches enable `stateful' logic operations via
  material implication,'' \emph{Nature}, vol. 464, pp. 873--876, 2010.

\end{thebibliography}
%
% <OR> manually copy in the resultant .bbl file
% set second argument of \begin to the number of references
% (used to reserve space for the reference number labels box)
%\begin{thebibliography}{1}

%\bibitem{IEEEhowto:kopka}
%H.~Kopka and P.~W. Daly, \emph{A Guide to {\LaTeX}}, 3rd~ed.\hskip 1em plus
%  0.5em minus 0.4em\relax Harlow, England: Addison-Wesley, 1999.
%\end{thebibliography}

%\end{thebibliography}

% that's all folks
\end{document}